\documentclass[aps,showkeys,twocolumn,preprintnumbers,notitlepage, secnumarabic,10pt, nofootinbib]{revtex4}
\usepackage{dcolumn}      % Align table columns on decimal point
\usepackage{bm}           % bold math
\usepackage[caption=false]{subfig}

\usepackage{graphicx}
\usepackage{xcolor}
\usepackage{rotating}
\usepackage{amsmath,amssymb}  % Better maths support & more symbols
\usepackage{bm}  % Define \bm{} to use bold math fonts
\begin{document}
\title{Dark Energy: A Dynamical Systems Approach to the Reconstruction of the Equation of State }
\author{Bob Osano$^{1,2}$ \\
{{$^{1}$Cosmology and Gravity Group, Department of Mathematics and Applied Mathematics, \\University of Cape Town (UCT), Rondebosch 7701, Cape Town, South Africa}}\\$\&$\\
{{$^{2}$Centre for Higher Education Development,\\ University of Cape Town (UCT), Rondebosch 7701, Cape Town, South Africa\\}}}
\date{\today}
\begin{abstract} This correspondence delves into the application of dynamical systems methodologies within the context of cosmology, specifically addressing a preliminary strategy for determining the range for the equation of state for dark energy ($\omega_{DE}$). Our findings suggest that the preferred range for $\omega_{DE}$ is between -1.1 and -0.6.
\end{abstract}

\maketitle
\section{\label{intro}Introduction} 
Determining the equation of the state of dark energy remains a challenge to cosmologists although it may appear as a foregone conclusion that this substance is responsible for the recent cosmic acceleration. Recent efforts in this regard include \cite{Rah,Col,Dai,Sin,Yang, Ave, Cam,Esc,Tri, Teng, Upa, Gon}. Whether the dark energy equation of state parameter \(\omega_{DE}\) is constant (\(\omega_{DE} \)) or evolving as described by \(\omega_{DE}(z) = \omega_{0} + \omega_{a} f(z)\) \cite{Col,Gon, Sin, Bob0, Bob1} remains uncertain. However, its nature and behaviour have profound implications for our understanding of the universe's evolution. We revisit this subject in this letter.

The letter is divided into two pertinent sections. The first section provides a comprehensive dynamical systems analysis approach to the matter-radiation transition. Although the detailed content in this section is standard and commonly found in existing literature, it offers foundational context for the methodology employed in the subsequent section. In the second section, we focus on analysing the transition from matter (M) to dark energy (DE). It is important to note that this approach traditionally presumes dark energy synonymous with the cosmological constant\cite{Ber}. 

In our investigation of dark energy, we shall forgo the assumption that it is synonymous with the cosmological constant. Instead, we propose a hypothetical scenario in which dark energy interacts with dark matter. Though much of the methodology presented here is derivable from forms extensively discussed in existing literature, an exhaustive recapitulation is unnecessary. However, for the sake of thoroughness, pertinent references will be provided, offering comprehensive guidance for the reader. We employ a dynamical systems approach, concentrating on identifying critical points where the flow alters its behaviour, rather than focusing on deriving explicit solutions to the equations of motion. The applicability of this methodology to cosmology has been demonstrated in \cite{Ber}.
 This letter brings it all together in an attempt to answer the question: Can dynamical systems techniques aid in the reconstruction of the EoS of dark energy? As this is a letter, we are deliberately brief. We commence our analysis by postulating a model comprising three constituents, designated as \( A \), \( B \), and \( C \), with the interaction occurring specifically between constituents \( A \) and \( B \). Additionally, we assume that the noninteracting component remains constant throughout the analysis. To connect with the existing literature, we will assume that the interacting components follow a barotropic form. Additionally, we will consider a flat cosmological model.
 The evolution timescale is quantified using the parameter \(\eta = \log a\), where \(a\) represents the scale factor. We utilise the prime notation to denote derivatives with respect to \(\eta\). Beginning with the Friedmann equations, we derive the continuity equations that describe the evolution of the energy densities of the individual constituents. These equations can be expressed in the following form:
\begin{eqnarray}\label{rhodots1}
{\rho}'_{A}&=&-3(1+\omega_{A})\rho_{A}+\frac{Q_{AB}}{H}\\\label{rhodots2}
{\rho}'_{B}&=&-3 (1+\omega_{B})\rho_{B}-\frac{Q_{AB}}{H}\\\label{rhodots3}
{\rho}'_{C}&=&0,
\end{eqnarray}
In this context, \( Q_{AB} \) denotes the interaction term between constituents \( A \) and \( B \). Subsequently, we will interpret this interaction as one occurring either between radiation and matter or between dark matter and dark energy. It is noteworthy that, in terms of the conservation law, the following relation holds:
\begin{eqnarray}
\nabla_{\mu} \left( \sum_{i=A,B} T^{\mu\nu b}_{i} \right) = 0,
\end{eqnarray}
where $\mu, \nu$ range over \( 1, 2, 3, 4 \). This indicates that the total energy-momentum tensor is conserved, but the energy-momentum tensors for each constituent part are not conserved independently.

\section{Expansion normalised formulation}
To establish a parallel between the conventional methodology and the novel approach presented in this letter, we adopt expansion-normalized variables akin to those utilised in \cite{Ber}. Within this framework, the components of the first Friedmann equation for a radiation and matter-dominated model are delineated as follows:
\begin{eqnarray}\label{norm1}
X&=&\Omega_{A}=\frac{\rho_{A}}{3H^2}\nonumber\\
Y&=&\Omega_{B}=\frac{\rho_{B}}{3H^2}\nonumber\\
Z&=&\Omega_{C}=\frac{\rho_{C}}{3H^2},
\end{eqnarray} and the first Friedmann equation takes the form
\begin{eqnarray}\label{fried1}
X+Y+Z=1
\end{eqnarray}The effective EoS for this system has the form
\begin{eqnarray}\label{eqeff}
\omega_{eff}=\bigg(\frac{\Sigma_{i}\omega_{i}\rho_{i}}{\Sigma_{i}\rho_{i}}\bigg),
\end{eqnarray}
where \(i=A, B, C\). 
It is important to note that we will adjust Equation (\ref{eqeff}) according to the predominant constituents during the specific epoch under investigation, which will be elaborated upon later in this letter. To maintain a more general framework, we will revert to the AB notation.
\section{A-B Dominated }
We present the evolution equations for A and B. It can be demonstrated that the evolution equations assume the following form.\begin{eqnarray} \label{xdot1}
X'&=&3X[-(1+\omega_{A})+(1+\omega_{A}+\frac{\alpha}{3})X+(1+\omega_{B})Y]
\nonumber
\\\label{ydot1}
Y'&=&3Y[-(1+\omega_{B})+(1+\omega_{A}-\frac{\alpha}{3})X+(1+\omega_{B})Y],
\nonumber\\
\end{eqnarray} By utilizing the constraint given by equation (\ref{fried1}), we have eliminated \(Z\) from these equations and have employed the ansatz \(Q_{AB} = \alpha H X\). The analogous form of equation (\ref{eqeff}) for this system can be expressed as follows: \begin{eqnarray}
 \omega_{eff}&=&-1+X(1+\omega_{A})+(1+\omega_{B})Y.
 \end{eqnarray}
We note that the system has the fixed points ($X^{*}, Y^{*}$)= (0,0), (0,1) and (1,0) regardless of the values of $\omega_{A}$ and $\omega_{B}$ when no interactions occur between $A$ and $B$. More important, is the fact that the EoS of $\omega_{A}$ and $\omega_{B}$ can be recovered from a generic portrait as will demonstrate.
\section{Matter-Radiation}
We know that \(\omega_{m} = 0\) and \(\omega_{r} = \frac{1}{3}\). However, the general form in equations (\ref{xdot1}) allows us to explore the scenario where the Equation of State (EoS) parameters for either component are unknown. Specifically, we can investigate whether it would be possible to recover these parameters. 

To relate this to existing literature, we consider the notation \(A \equiv r\) (radiation), \(B \equiv m\) (matter), and \(Q_{AB}\equiv Q_{rm} \) (the case of no interaction). Under these conditions, we have:\begin{eqnarray} 
\label{xdot2}
X'&=&X(-3+(3+\alpha)X+4Y)\nonumber\\\label{ydot2}
Y'&=&Y(-4+(3-\alpha)X+4Y).
\end{eqnarray}  We have retained the case where \(\alpha \ne 0\) to examine whether any interactions are present. We will later set \(\alpha = 0\) to align with existing literature. The corresponding equation to (\ref{eqeff}) for this system is:\begin{eqnarray}\label{constr1}
\omega_{eff} = -1 + X+ \frac{4}{3}Y.
\end{eqnarray}
The fixed points for the system described by equation (\ref{ydot2}) are \((0,0)\), \((0,1)\), and \(\left(3/(3+\alpha), 0\right)\). By setting \(\alpha = 0\), these fixed points can be interpreted as representing the dark-energy, radiation, and matter-dominated epochs, respectively \cite{Ber}. We will denote these points as \(R(0,1)\), \(M(1,0)\)and \(D(0,0)\) . These points form a triangle that we will henceforth refer to as the RMD- triangle.

\begin{figure}[htb]
\caption{Radiation- Matter: Phase portraits $X$ (horizontal axis) and $Y$ (vertical axis), with varying $\alpha$.
 \protect\subref{subfigaa} $ \omega_{A}\equiv\omega_{r}=1/3$,$\omega_{B}=\omega_{m}=0$. 
 \protect\subref{subfigbb} $ \omega_{A}\equiv\omega_{r}=1/3$, $\omega_{B}=\omega_{m}=0$, $\alpha\ne0$. 
 \protect\subref{subfigcc}$ \omega_{A}\equiv\omega_{r}=1/3, \omega_{B}=1\ne\omega_{m}$, $\alpha=0$. 
\protect\subref{subfigdd}$ \omega_{A}\equiv1\ne\omega_{r}$, $\omega_{B}=\omega_{m}=0$, $\alpha=0$.
}
\label{fig:aa}
\subfloat[\label{subfigaa}]{\includegraphics[width=0.45\columnwidth]{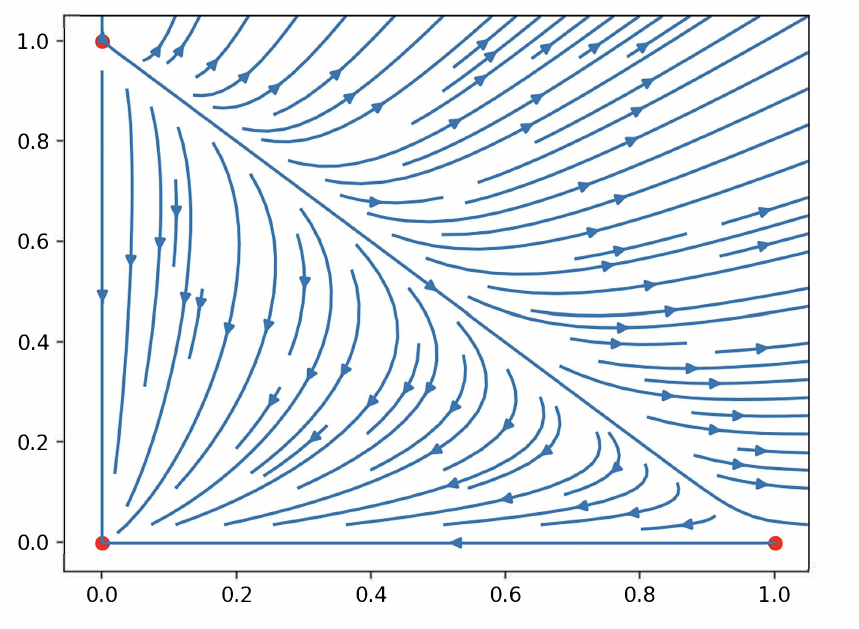}}
\subfloat[\label{subfigbb}]{\includegraphics[width=0.43\columnwidth]{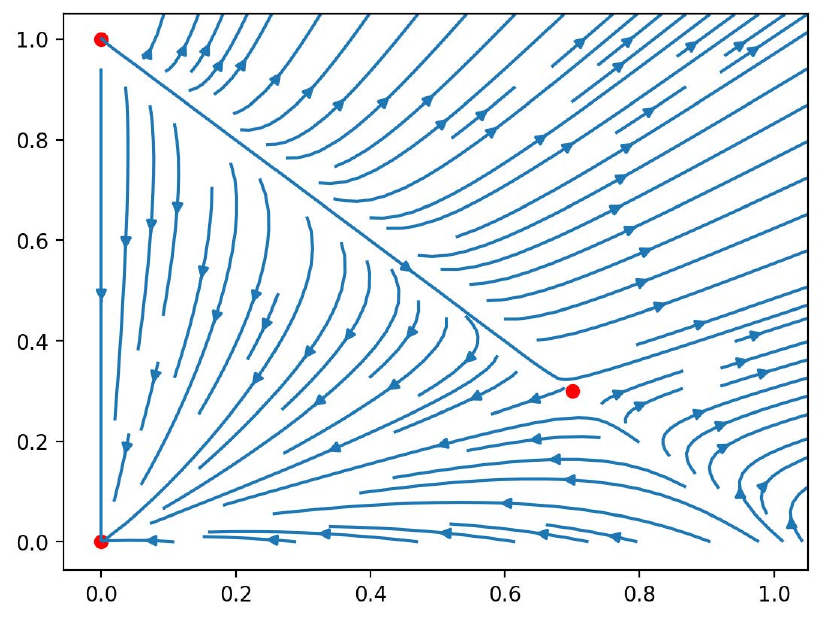}}\\
\subfloat[\label{subfigcc}]{\includegraphics[width=0.45\columnwidth]{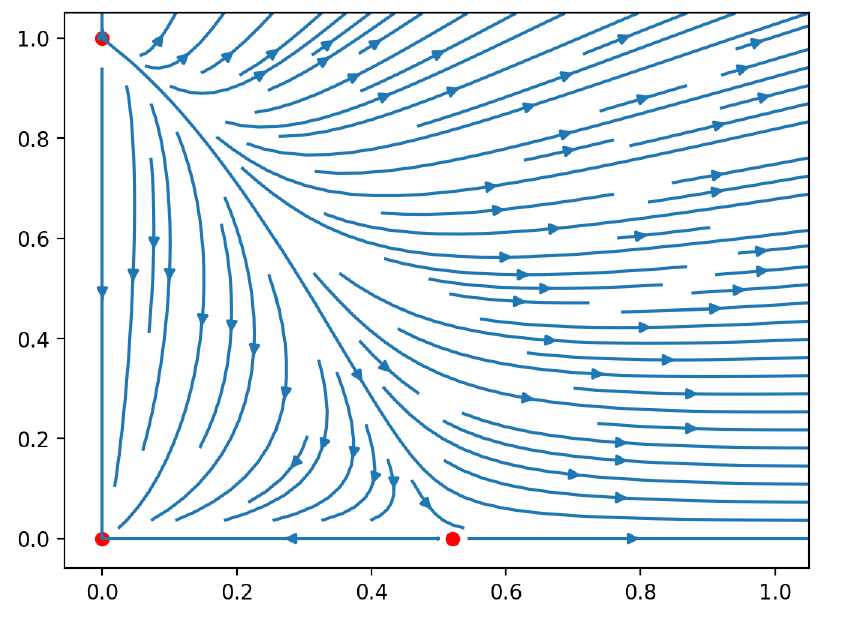}}
\subfloat[\label{subfigdd}]{\includegraphics[width=0.45\columnwidth]{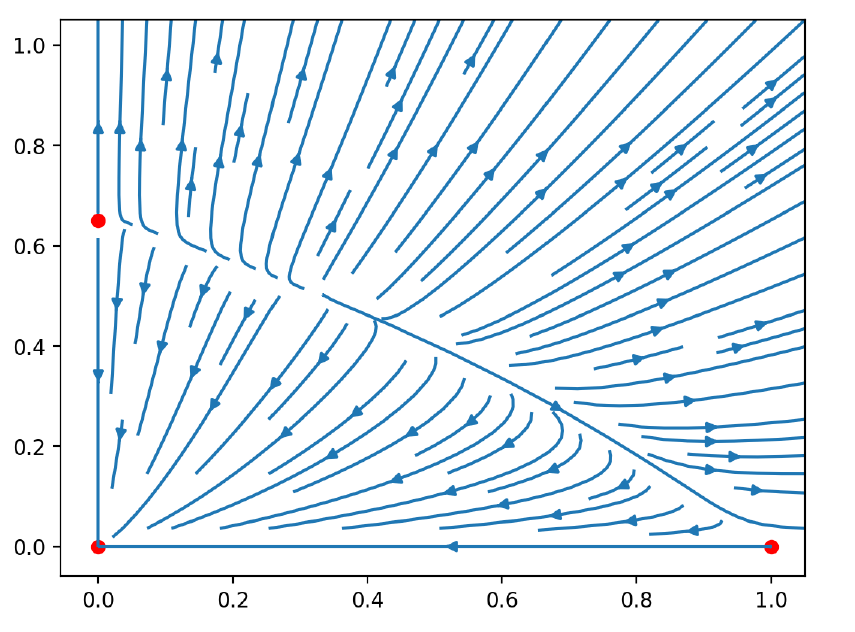}}
\end{figure}

The phase portraits for this system are shown in Figures (\ref{fig:aa}a-\ref{fig:aa}d). 
It is possible that when \(\omega_{eff}\) is known for each fixed point, the corresponding EoS parameters for radiation and matter can be determined straightforwardly and unambiguously.

Figure (\ref{fig:aa}a) depicts the standard portrait with parameters \(\omega_{r}=1/3\), \(\omega_{m}=0\), and \(\alpha=0\). In this figure, the region above the line \(X+Y=1\) is deemed non-viable\cite{Ber}. Figures (\ref{fig:aa}a-\ref{fig:aa}d) present generic cases where one of the constituents corresponds to either radiation or matter. It can be shown that if the value of \(\omega_{eff}\) is known, for instance through observational data, one can reconstruct the Equation of State (EoS) of either component by utilizing the node values from the phase portraits. The first thing to note is that a non-zero value for the interaction term moves one of the nodes along the hypotenuse of the RMD triangle, while a different value of EoS moves a node along the horizontal or vertical edges of the RMD triangle.

Consider Figure (\ref{fig:aa}c), where the nodes are located at \(D(0,0)\), \(R(0,1)\), and \(M(0.5,0)\). The final node indicates that \(X=0.5\) while \(Y=0\). Given that the horizontal arm is matter-dominated, we deduce \(\omega_{eff}=0\). Using this information along with equation (\ref{constr1}), we can reconstruct \(\omega_{B}\) and obtain the value \(\omega_{B}=1\). Similarly, it can be demonstrated that \(\omega_{A}=1\) for the point situated in the radiation-dominated arm. This rudimentary approach provides an initial estimate for reconstructing a given Equation of State (EoS) from such phase portraits. For a rigorous analysis, independent determination of \(\omega_{eff}\) is essential. As will be discussed in the subsequent section, \(\omega_{eff}\) can be derived from observational data and its link to the second Friedmann equation.

\section{Matter - Dark Energy }
Let us now use the knowledge from the previous section to attempt to construct the EoS of dark energy. In this case, we let $\omega_{A}\equiv\omega_{DM}$ and $\omega_{B}\equiv\omega_{DE}$. it follows that
\begin{eqnarray} \label{xdot3}
X'&=&3X[-1+(1+\frac{\alpha}{3})X+(1+\omega_{DE})Y]
\nonumber
\\\label{ydot3}
Y'&=&3Y[-(1+\omega_{DE})+(1-\frac{\alpha}{3})X+(1+\omega_{DE})Y],
\nonumber\\
\end{eqnarray} where \(Q_{AB} = \alpha H X\). With $\omega_{DM}=0$, the equivalent of equation (\ref{eqeff}) for this system assumes the form \begin{eqnarray}\label{eff1}
 \omega_{eff}&=&-1+X+(1+\omega_{DE})Y.
 \end{eqnarray} Note that the fractions \( X \) and \( Y \) are functions of redshift \( z \), given that they are normalized with respect to the expansion as shown in Equation (\ref{norm1}). Phenomenologically, we can express the last two terms as \(\omega_{DDE} f(z)\), where \(\omega_{DDE}\) is the dynamical equation of state that accounts for both matter and dark energy. The function \( f(z) \) could take a linear, logarithmic, or hyperbolic form \cite{Bar}. The interaction mediating these two components allows us to extend the definition in this manner. If \(\omega_0 = -1\), then equation (\ref{eff1}) becomes
  \begin{eqnarray}\label{eff2}
 \omega_{eff}&=&\omega_{0}+\omega_{DDE}f(z).
 \end{eqnarray}
 Equation (\ref{eff2}) has the form of the EoS often used for dynamical dark energy. We emphasise that the difference, compared to what is in literature \cite{CPL} is that our dynamical part incorporates matter. We can now establish a relationship between the deceleration parameter, \( q \), and the variables \( X \) and \( Y \) by utilizing both Friedmann equations, yielding the following expression:
 \begin{eqnarray}
\frac{{H'}}{H}&=-(1+q)=&-\frac{3}{2}\bigg[X+(1+\omega_{DE})Y\bigg].
 \end{eqnarray} For the standard \(\Lambda\)CDM model, the deceleration parameter determined from local observations is \( q_{0} = -0.55 \). Other observations may yield slightly different values. Generally, the value of \( q_{0} \) from various observations tends to fall within the range \(-0.8 \leq q_{0} \leq -0.4\). We will use this range to illustrate our approach.
 
\begin{figure}[htb]
\caption{Matter- Dark Energy: In these portraits $X$ is the horizontal axis and $Y$ the vertical axis. 
\protect\subref{fig:subfiga}$\omega_{B}\equiv\omega_{DE}=-1.1$, $\omega_{A}\equiv \omega_{M}=0$, $\alpha=0$. 
\protect\subref{fig:subfigb} $\omega_{B}=-0.6$, $\omega_{A}\equiv \omega_{M}=0$, $\alpha=0$.
\protect\subref{fig:subfigc} $\omega_{B}=-0.6$, $\omega_{A}\equiv \omega_{M}=0$, $\alpha=0.8$.
\protect\subref{fig:subfigd}$\omega_{B}\equiv\omega_{DE}=-1.1$, $\omega_{A}\equiv \omega_{M}=0$, $\alpha=0.8$.
}\label{fig:b}

\subfloat[\label{fig:subfiga}]{%
  \includegraphics[width=0.44\columnwidth]{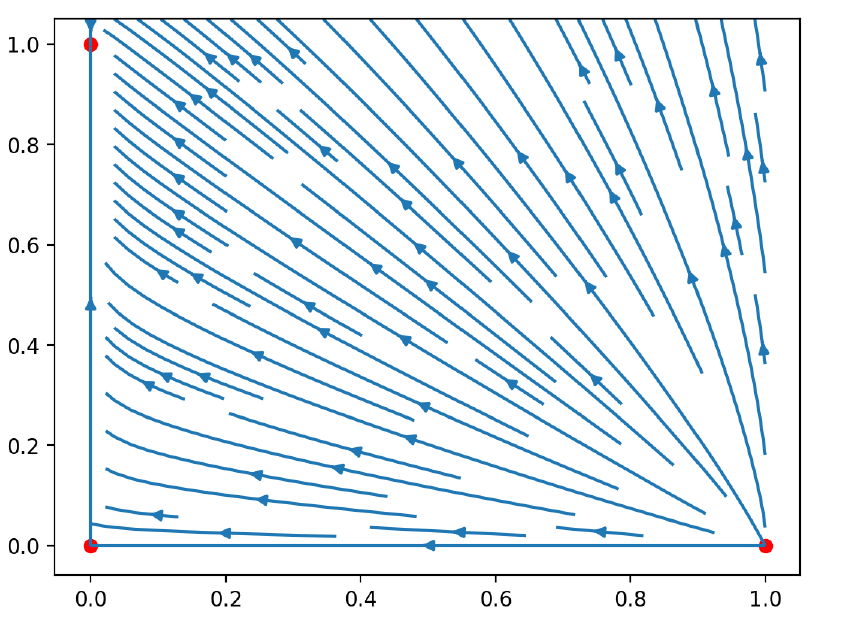}%
}
\subfloat[\label{fig:subfigb}]{%
  \includegraphics[width=0.47\columnwidth]{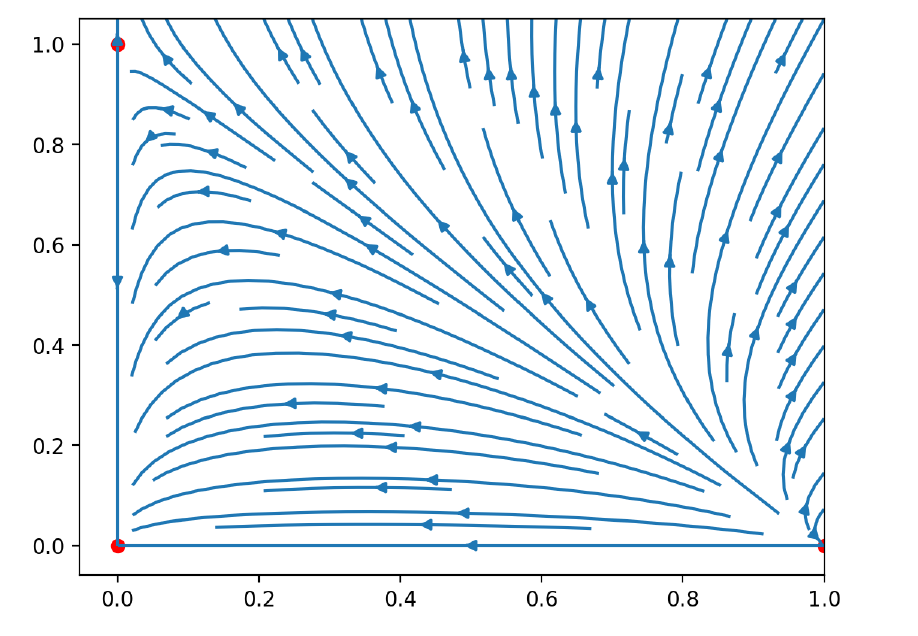}%
}\\
\subfloat[\label{fig:subfigc}]{%
  \includegraphics[width=0.45\columnwidth]{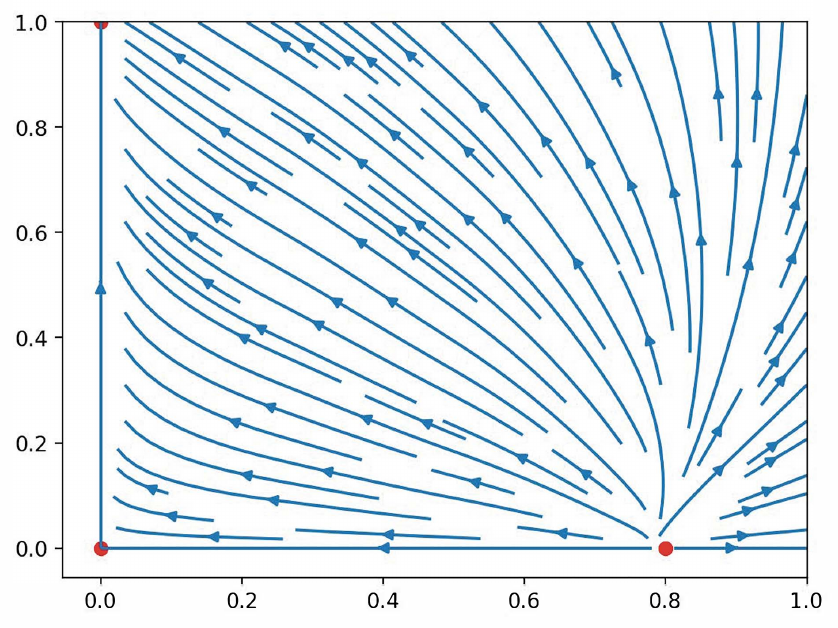}%
}
\subfloat[\label{fig:subfigd}]{%
  \includegraphics[width=0.45\columnwidth]{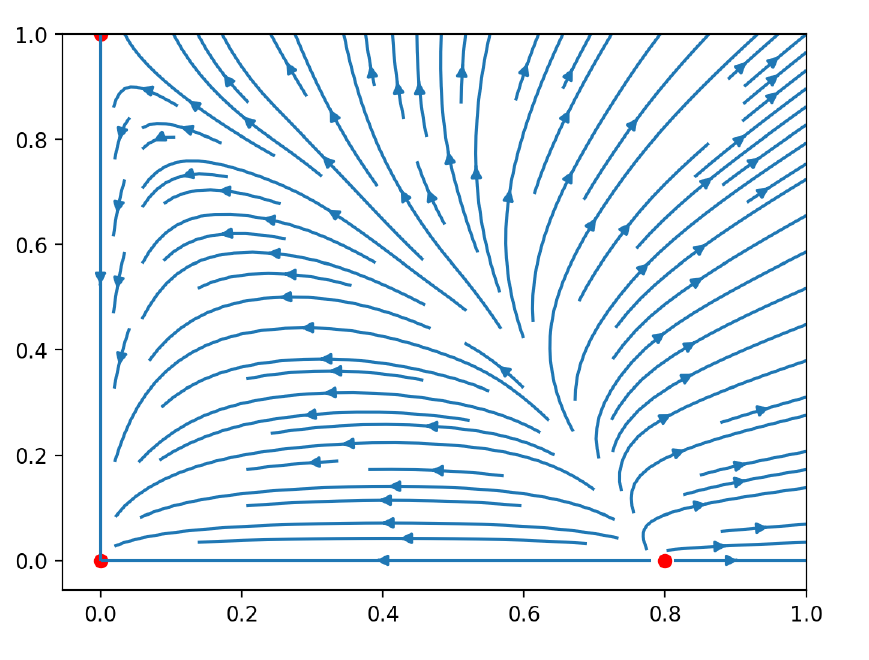}%
}

\end{figure}

We considered the range \(-1.1 \leq \omega_{DE} \leq -0.6\) based on the implication of \(q_0 = -0.6 \pm 0.2\) representing the average of several experiments with the sweet spot being \(q_0 = -0.55\)\cite{Cam, Nai, Muk}. The phase portraits indicate that, for \(\omega_{DE} = -1.1\), the node \((1,0)\) is the future attractor, while for \(\omega_{DE} = -0.6\), the node \((0,0)\) serves as the future attractor. A non-zero interaction acts as a bifurcation parameter, shifting one of the nodes horizontally or vertically. 
\section{Conclusion}
This letter explores the question of whether it is possible to reconstruct the equation of state (EoS) for one of two competing constituents when the effective EoS is known or can be determined experimentally. We conclude that this is indeed feasible. Employing dynamical techniques, we have demonstrated the possibility of reconstructing the EoS in a model consisting of both matter and radiation. Furthermore, we have applied the same methodology to a model comprising dark energy and dark matter, confirming the applicability of the technique in this context as well. It is important to note that this approach requires prior knowledge of the effective equation of state, which can be derived from observational data. Our findings indicate that \(-1.1 \leq \omega_{DE} \leq -0.6\). In the specified range, the lower value results in a future attractor node located at \((0,1)\), while the upper value designates \((0,0)\) as the future attractor, consistent with the presence of a cosmological constant. If the former case is confirmed, it could negate the cosmological constant being the long-sought-after dark energy (DE). This finding would also confirm the crossing of the cosmological constant boundary \cite{Upa, Sta} and point towards the "Big Rip" \cite{Cal} as a potential fate of the universe unless a new and yet undetermined form of energy intervenes.

It is essential to underscore that the results and interpretations are contingent upon the precision of the deceleration parameter measurements. Despite its rudimentary nature, this approach offers a foundational basis for subsequent inquiries and in-depth analyses. The dependence of the equation of state (EoS) parameter on the value of the Hubble parameter, coupled with the increasing discordance in its measurement across various methodologies, underscores the need for further investigation into dynamic dark energy EoS models as well. Such investigation is essential to elucidate the relationship between the expansion rate and the evolution of dark energy.

\end{document}